# High Density Through Silicon Via (TSV)

Mr. Magnus Rimskog, M.Sc. and Mr Tomas Bauer, M.Sc.
Silex Microsystems
Bruttovagen 3
SE-175 26 Jarfalla, SWEDEN

*Abstract*- The Through Silicon Via (TSV) process developed by Silex provides down to 30 µm pitch for through wafer connections in up to 600 µm thick substrates. Integrated with MEMS designs it enables significantly reduced die size and true "Wafer Level Packaging" - features that are particularly important in consumer market applications. The TSV technology also enables integration of advanced interconnect functions in optical MEMS, sensors and microfluidic devices. In addition the via technology opens for very interesting possibilities considering integration with CMOS processing. With several companies using the process already today, qualified volume manufacturing in place and a line-up of potential users, the process is becoming a standard in the MEMS industry. We provide a introduction to the via formation process and also present some on the novel solutions made available by the technology.

## 1 INTRODUCTION AND BACKGROUND

This presented via technology was developed in an environment consisting of a fully equipped 1000 m² state-of-the-art 6" wafer fab. With regards to manufacturability, cost efficiency and simplified back-end, the via technology has great benefits. Packaging for example is often considered a limiting factor when quickly taking a product from design concepts to working prototypes and the following volume manufacturing. Following the general trend, there is a need to continuously work on developing proprietary MEMS foundry processes, leveraging on intellectual property and know-how to complement the standard manufacturing capability. The via technology is an excellent example of such MEMS process enabling true wafer level packaging and MEMS designs with significantly reduced form factor. The technology has already been applied to a diverse range of products, and while not offering any own products to the market, a MEMS foundry is in a unique position to promote the benefit of the technology to a wide range of applications. The via technology as presented here is applied to a number of different types of products from low volume applications to high volume consumer electronic markets.

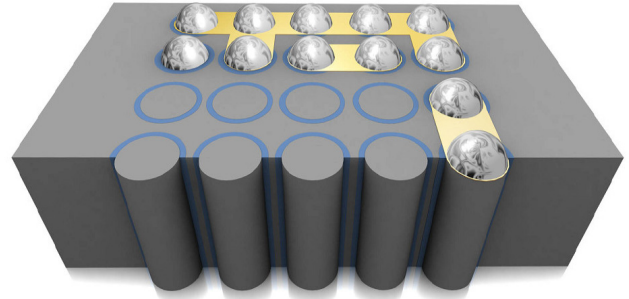

*Figure 1: Schematic cross-section of through wafer interconnects.*

## 2 TECHNOLOGY

In response to demand for through silicon interconnect functionality a novel idea to isolate a section of a low resistivity silicon wafer laterally by incorporating a trench filled with an isolating material [1] was invented. This means that the actual via plug is there from beginning and the difficulties with homogenous fill of a wide plug area is avoided. The isolating trench can have various shapes however most often it is shaped as a circle to minimize any stress. The trench can also take other shapes if necessary as long as it constitutes a closed loop in order to avoid shorting to the base substrate. The process uses a DRIE process in order to make a trench. This DRIE process has been developed in order to achieve the necessary high aspect ratio features in up to 600 µm thick substrates but more important it has been optimized in order to provide a fast etch time. The trench width is varied depending on actual design criteria with typical width in the order of 10 to 15 µm.

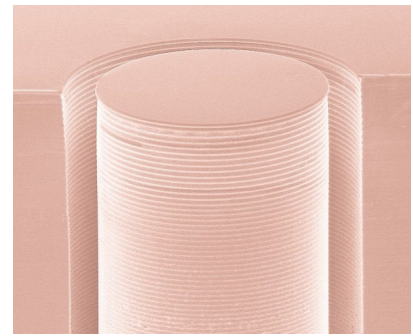

*Figure 2: SEM picture of through wafer via "plug", formed using DRIE.*

In order to keep the via plug in place, the trench etch stops short of reaching all the way through the wafer. Following this etch, the wafer is subject to a filling process, which fills up the trenches with a dielectric material providing necessary isolation from the remaining bulk. In several customer applications the via has been proven to be able to hold a good vacuum level and cross section analysis shows that with the qualified via fill process voids can be avoided. As a final step, a CMP process is applied to the backside of the wafer, removing the material that keeps the "via plugs" connected to the bulk, thereby isolating the via plugs from the bulk of the wafer.

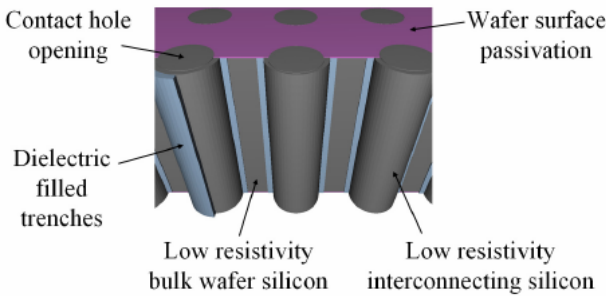

*Figure 3: Schematic cross section of via substrate features.*

In some sense the via technology could be compared with the Silicon on Insulator (SOI) technology, noting the 90° difference in the orientation of the buried insulator versus the plane of the wafer. It is also possible to integrate the via technology into SOI material hence enabling further interesting option with regards to continued processing.

## 2.1 Typical Design Flow

Detailed design rules can unfortunately not be detailed in this paper, but some general guidelines can be given with respect to integrating via technology on wafer level. The typical procedure begins with the outline of the requested position of the through wafer connections on the wafer. For smallest possible pitch between vias, down to 50 μm or less, it is necessary to have neighboring vias share the same isolating trench. Since actually a whole wafer without any processing can be considered as a through wafer connection there is no upper limit in terms of the size of the isolated area. As the designer can choose the trench layout freely, assuming a uniform trench width is used over the entire wafer, and the resistance is proportional to the cross section of the connecting silicon, shapes can be chosen that efficiently utilize the available space. Shapes are in many cases optimized towards providing a sensing or an actuating function towards the device silicon in the design. Often there is a requirement for a tight pitch in only one axis of the two-dimensional space and less of a limitation in the other axis. One example of this is the contact pads surrounding a CMOS chip. In such case, a larger total cross section area of the via connection, and thereby lower resistance, can be achieved by outlining the vias in rectangular shape. By using the available chip area to a maximum, while maintaining a small pitch along the other axis resistance can be minimized. Following customer input on the design layout, engineers performs a cross check against design rules, taking into consideration wafer thickness and structural stability versus possible post processing requirements.

## 2.2 Process Implementation

From a technological standpoint, the described via process is primarily a "via-first" process as it needs a high temperature step during filling. From a MEMS processing perspective, the "via-first" approach is in most cases very suitable as in most often there is a desire to minimize additional wafer level post processing and handling of for example released MEMS structures since these can be rather delicate. As mentioned, the completed via substrate wafer resembles the SOI wafer as it allows more or less unlimited processing of the substrate, even at temperatures exceeding 1000°C. This means the via substrate wafer will be able to enter the MEMS device manufacturing flow as part of the starting material.

Integrating the via technology with traditional SOI technology enables combining the technology with applications requiring a high resistivity device layer. The original via process requires a low resistivity bulk part to create a low loss electrical connection, but by integrating the via process into an SOI wafer with a low resistivity handle and high resistivity device layer this can still be achieved. Depending on the post processing requirements, the contacting to the via sections in the handle part of the substrate is made by one of the existing conductive layers or alternatively by incorporating shallow plugs of doped polysilicon into the design.

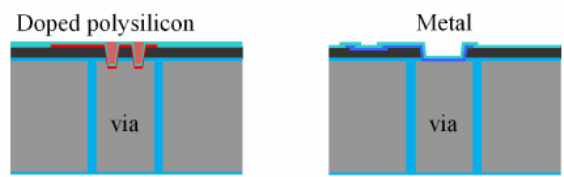

*Figure 4: Integration of via process in SOI wafer with high resistivity device layer.*

Lately a variation of the via technology involving localized doping allows also for low resistance connections through bulk material with higher resistivity. This, as well as the combination with SOI material as described above increases the possibility for integration to a wider amount of designs and applications.

## 3 IMPLEMENTATION

When debating pros and cons of MEMS technology, not rarely has the argument brought forth that the main hurdles relating to commercial implementation is packaging and interconnect. With the increasing maturity of the MEMS industry a number of special solutions has been brought forward and are also starting to turn into standard solutions.

With the successful integration of the via technology described herein to more than 10 various products over a large area of application fields the via technology offered provides one such standard. In short a well qualified via process virtually takes care of one of the inherent, most significant problems associated with MEMS by providing true wafer level packaging of MEMS. Depending on the designs in question the components can either be mounted directly onto PCB or combined with ceramic packaging technology.

### 3.1 Cost Factor

The via substrate manufacturing flow, containing a through wafer DRIE process and the following filling and CMP, does add on manufacturing cost. In volume production this cost is comparable to an overall cost similar to that of an SOI wafer. The reduced form factor and wafer level packaging of the MEMS die, enabled by the technology, makes the implementation beneficial for most applications and is especially effective with small die sizes where bond pad area takes up a significant part of the die size. It also in several cases constitutes an enabling technology which leads to new market opportunities with regards to component development.

### 3.2 Implementation Examples

To date, the via process has been offered to a number of foundry customers as an added value building block for integration in customer specific designs. One of the most unmistakable examples of via process implementation is wafer level packaging of MEMS devices. Metal routing and contact pad area in many cases consumes a large portion of many MEMS die layouts. Adding a standard ceramic package, the final component will in many cases be ten times larger in area than the active die itself. Incorporating via technology for interconnect and wafer level packaging significantly reduces the form factor while getting simplifying a large share of the traditional "back-end" process.

Since the via substrate allows for high temperature processing fusion bond can in many examples be used for the capping process. This enables seal ring size to be reduced adding further to the increase in number of dies per wafer. Also the technology solves any issues with exposing bond pads through elaborate two step dicing methods. The dies can be singulated with a single cut by a dicing saw and are surface mountable straight of the dicing line.

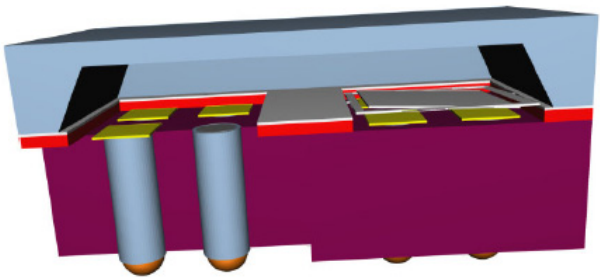

*Figure 5: Example of wafer level packaging of MEMS device with 3D interconnect by Silex silicon Vias.*

Another area of interest for implementation is the integration of MEMS and CMOS. Via technology provides new possibilities for such integration although yield factors for particular designs needs to be carefully considered when integration method (wafer level or die level) is chosen.

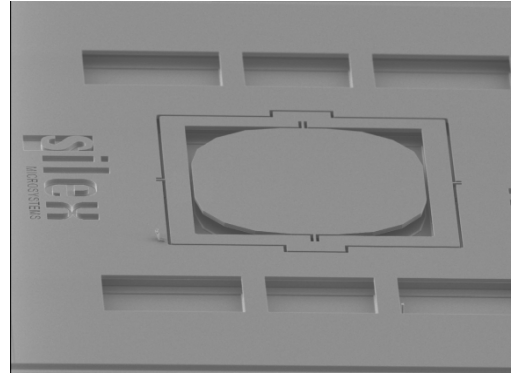

*Figure 6: Single element MEMS micromirror with integrated Vias.*

For many sensing devices such as flow- and pressure sensors the via technology simplifies packaging and connection to the chip. The connection to the sensing chip can be made at the chip side facing away from the media while having the sensing part of the die in direct contact with the measured media.

In cases of dies made up of large arrays of "facets" extensive metal routing in the plane of the active area is often required. By using through wafer via technology, real estate can be saved while improving fill factor. Metal routing and flip chip mounting is accommodated on the back side of the die beneficial for example in imaging applications.

Using via technology in micro-fluidic devices enables creation of electrodes in the channels, and at the same time it allows the integration of more additional MEMS features such as fluidic filters, high aspect ratio structured channels and silicon to glass bonding.

The via technology is also applicable when using the System in Package (SiP) approach, integrating a number of integrated circuits in a single module. The SiP can perform many of the functions of an electronic device, such as a mobile phone. This feature is particularly valuable in space constrained environments like mobile phones as it reduces the form factor and complexity of the PCB and overall design.

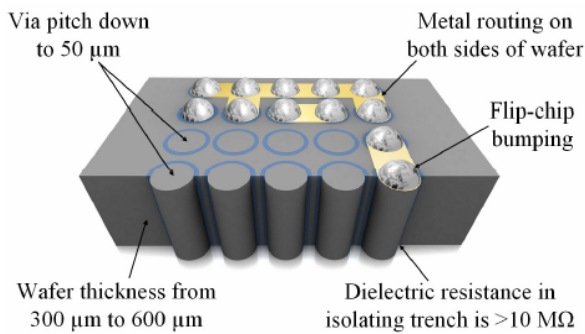

*Figure 7: Interposer solution for system in package (SiP).*

There is also a continued drive for advanced small scale packaging, which can benefit from use of other materials than ceramics to achieve the accuracy and reduction in footprint requested. Traditional packaging based on low temperature or high temperature co-fired ceramics (LTCC or HTCC) are reaching limitations when using existing manufacturing methods. These technologies have difficulties achieving the small pitch for the electrical feedthroughs (<100 μm) and tight tolerances are required when moving closer to "micro scale". In that respect, silicon provides a good candidate with its long term reliability and high precision micromachining capabilities. In this respect, via technology and MEMS processing technology can be implemented as an alternative to realize customized silicon based wafer level packaging solutions.

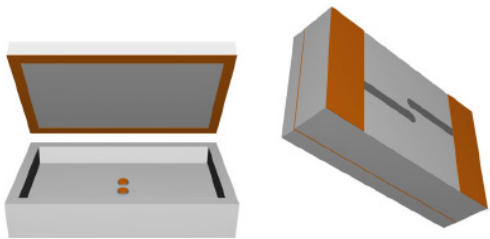

Figure 8: Wafer level micro scale packaging enabled by Silex via technology.

## 4 TECHNOLOGY LIMITATIONS

Current limitations when implementing silicon via technology are typically found in the domain of RF applications. When gold signal lines are not good enough as in many high frequency circuits and solutions are sought in copper, the silicon via will not do the job. Typical resistance in a standard via connection of 100 μm diameter in a 400 μm thick substrate is in the order of 1 Ohm and such resistance will in most RF designs constitute a loss higher than acceptable. There is extensive work in many places striving to enable a hermetic via solution with lower resistance but that is rather a subject for separate articles. No doubt such a technology proven in volume will enable even further development of MEMS and IC technologies while having limitations with regards to post processing allowed. A void free metal interconnect technology will for sure complement the Silicon via technology when further reduction of interconnect resistance is necessary.

## 5 COMPETING TECHNOLOGIES

via technologies is not a new concept in the MEMS or in the IC community and are seeing increased success in terms of commercial viability. However, the various approaches are still to a large extent struggling with inherent technological difficulties tightly linked to the chosen technology. Typical issues with metal based Vias have been the thermal mismatch with silicon causing micro cracks and reliability issues. Furthermore, a metal via must be incorporated towards the end of a MEMS process as the thermal budget of would not allow too much of high temperature post processing.

## 6 TECHNOLOGY LICENSING

With several foundry customers using the process today and an extraordinary line-up of potential users, it is of great interest to facilitate making this process a standard in the MEMS industry. In this context, the via technology is offered through licensing agreements and technology transfer programs to selected customers and partners who would favor to incorporate the technology in their manufacturing lines [1].

## 7 CONCLUSIONS

This paper has detailed information on a commercially available through wafer via process that provides a viable solution to some inherent problems associated with MEMS (packaging and interconnect). It has also provided examples of some of the many features made possible through this new technology. The via process has now been in production since 2005 for applications ranging from advanced medical devices to mobile phones, and since mid 2006, the process is offered as a standard MEMS foundry process.

To conclude this paper, key features characterizing this through silicon via process are listed below:

- High density via technology (pitch <50 μm)
- "No metal" starting material with unrestricted MEMS or CMOS post processing capability
- Enables true wafer level packaging of MEMS devices with vacuum sealing
- Low cost solution due to minimized component form factor and "all silicon package" for SMD assembly
- First through wafer via technology in volume production